# Miniature Neutron Spectrometer for Space

**C. Potiriadis[a], I. Kazas[b], C. Papadimitropoulos[a] and C. P. Lambropoulos [c,d][1]**

[a] *Greek Atomic Energy Commission,*
*Patriarxou Grigoriou & Neapoleos, Agia Paraskevi- Attiki, 15310 Greece*

[b] *Institute of Nuclear and Particle Physics, National Center for Scientific Research Demokritos,*
*Patriarxou Grigoriou & Neapoleos, Agia Paraskevi- Attiki, 15310 Greece*

[c] *ADVEOS Microelectronic systems P.C., Chalandri - Attiki, 15231, Greece.*

[d] *National and Kapodistrian University of Athens, Athens,Greece.*
*E-mail:* `lambrop@uoa.gr`

ABSTRACT: MIDAS is a miniature detector developed with purpose to assess the radiation field parameters near to an astronaut. Part of the device is a spectrometer for fast neutrons. In missions outside the geomagnetic field, fast neutrons are secondary products of the interaction of Galactic Cosmic Ray heavy ions with the materials in the spaceship or even the astronaut body. The Relative Biological Effectiveness of fast neutrons is high. The neutron spectrometer first prototype has been developed, calibrated and used for measuring $^{252}$Cf spectra.

## Contents

---

[1]Corresponding author.

## 1. Introduction

The European Space Agency (ESA) requires a miniature device sensitive to both charged particles and neutrons to enable measurement of dose, dose rate, energy deposition, linear energy transfer (LET) and calculation of dose equivalent. The goal is to increase the crew autonomy as far as it concerns operational decisions related to radiation hazards, mainly in missions outside the geomagnetic field. For neutrons the energy range sought is from 100 keV up to 200 MeV. The relative biological effectiveness of neutrons is high and the weighting factors for the calculation of dose equivalent are between 5 and 21 in the energy region between 100 keV and 100 MeV [1]. The device mass limit is put at 50 g and the dimensions should be less than 5 x 5 x 1 cm$^3$. Power consumption should allow 30 days of autonomous operation. The response to these requirements is the MIDAS device. A brief overview will be given in the first section and then the neutron measurement part of the device will be presented.

## 2. MIDAS description

The first prototype of the device can be seen in Figure 1(a). The conceptual design of the current prototype can be seen in Figure 1(b). The core of the device is the "sensitive cube". It consists of a plastic scintillator with neutron/gamma discrimination capability enclosed in a titanium box 1 mm thick. The scintillator dimensions are 7 x 7 x 7 mm$^3$ and its bottom face is connected to a silicon photomultiplier (SiPM) to collect the light signal. The five of the six external surfaces of the titanium box are covered by two layers of silicon pixel sensors. The aluminum enclosure is 1 mm thick, except from a circular area on the top and bottom cover with 0.5 mm thickness, with purpose to let 10 MeV protons reach the detector.

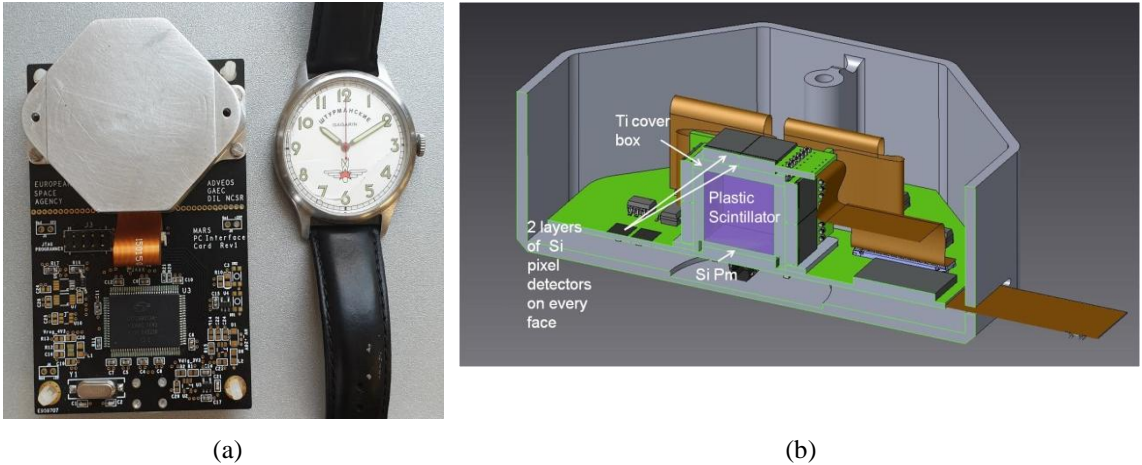

(a)          (b)

**Figure 1.** (a) The first prototype of the detecting head mounted on a custom data acquisition board. (b) Cross section view of the conceptual design of the second prototype.

When a fast neutron reaches the plastic scintillator, it has some probability to scatter elastically with the hydrogen nuclei. The recoil proton will produce a light signal and if its kinetic energy is less than 18 MeV, it will be absorbed by the titanium box and it will not produce signal in the surrounding silicon detectors. When a charged particle crosses the detector it will produce signal to the silicon detectors and light in the scintillator. The silicon detectors can provide for each event the energy deposited in the silicon pixels affected and their address. The silicon detectors are depleted monolithic active pixel sensors (DMAPS), which are designed for the first time to measure energy depositions from Galactic Cosmic Rays and Solar Energetic Particles.

## 3. The MIDAS' neutrons measurement system

The plastic scintillator EJ299-33 and its successor EJ276 of Eljen Technology are used. In the one out of the six facets of the scintillator is attached the MicroFC-60035-SMT Silicon Photomultiplier (SiPM) by ON Semiconductor (formerly SensL). The SiPM has an active area of 6 x 6 mm$^2$ and 18980 pixels. Every pixel of the SiPM acts as a Geiger Muller avalanche counter. The readout is based on discrete off the shelf components and it is coordinated by an IGLOO-e_AGL250_CS196 FPGA by MicroSemi, which is the controller of the whole sensitive cube. The readout circuit consists of a transimpedance amplifier which receives the current output of the SiPM. The output of the amplifier feeds a comparator and an integrator. The integrator is normally at reset state, but when a current pulse is produced in the SiPM the comparator triggers the integrator to leave the reset state and start integrating the output of the transimpedance amplifier. This integral is sampled twice by a very low power analog to digital converter: At first 0.5 μs after the comparator's firing (S1) and second time after 6 μs (S2). The pulse provided by the transimpedance amplifier is stretched in time in order to relax the speed requirements and thus reduce the power consumption. A result of this is that the dark pulses overlap the one with the other and induce fluctuations of the integrated output. It can be seen in Figure 2(c) and 2(d) that the spectrum taken with the aid of a $^{57}$Co source is clearly separated by the spectrum of fluctuations of the baseline and that the fluctuation of the baseline increases with temperature.

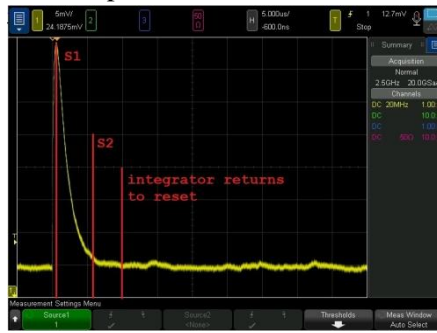
(a)

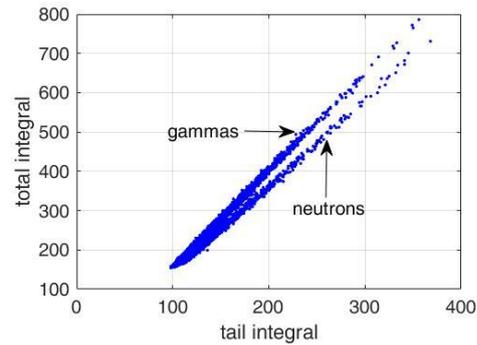
(b)

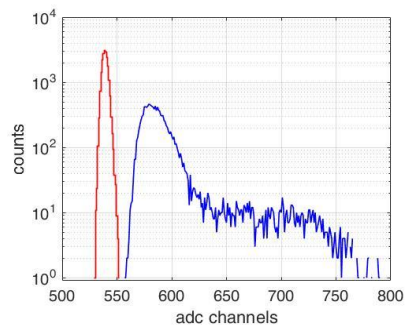
(c)

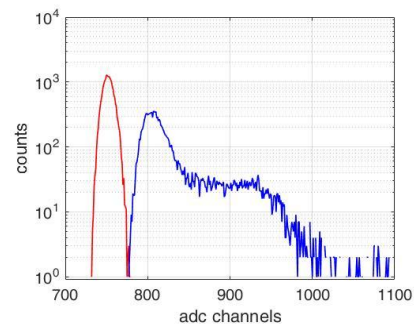
(d)

**Figure 2.** (a) Oscilloscope picture of the output of the transimpedance amplifier. This signal is integrated and the integral Si is sampled two times. (b) The scatter plot with x axis the S2-S1 and y axis the S2 is plotted and two branches appear. The branch with higher S2-S1 for the same S2 is due to neutrons. At low values the two branches cannot be distinguished. (c) The spectrum without source (red line) compared to the spectrum with a $^{57}$Co source (blue line) at 0$^o$C. (d) The same spectra recorded at 27$^o$C.

Pulse Shape Discrimination (PSD) is a method for discriminating fast neutrons from gamma rays based on the fact that the scintillation light consists of a slower decay component compared to the gamma rays. Inside the scintillator, the protons produced from neutron interactions exhibit larger energy transfers and shorter ranges. Short range increases the probability of a special case of Dexter energy transfer: the triplet–triplet annihilation which results in delayed fluorescence [2]. Gamma ray – neutron discrimination is achieved by introducing two integral windows per pulse, the total integral (S2) and the tail integral (S2-S1). If these integrals are displayed in a scatter plot, as shown in Figure 2(b), a clear separation of neutrons from gamma rays can be seen.

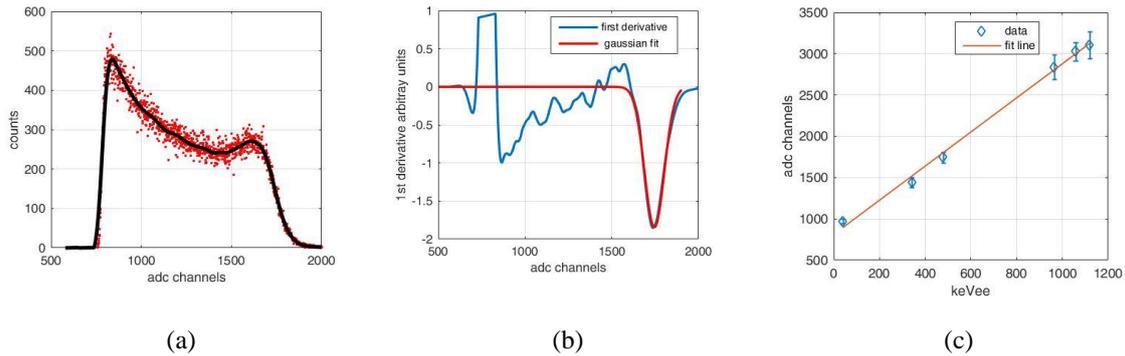

(a) (b) (c)

**Figure 3.** (a) The experimental spectrum recorded with a $^{137}$Cs source (red dots) and the smoothed spectrum (black line). (b) The first derivative of the smoothed spectrum and the Gaussian distribution at the region of interest. (c) The calibration line which relates the deposited electron energy in the plastic scinitllator to adc channels.

### 3.1 Calibration with electron energy deposition

As the plastic scintillator consists of materials with low atomic number the probability of electron production through photo-conversion is very low, while the dominant interaction is Compton scattering. Consequently the conversion of the analog to digital converter channels to electron energy is based on the determination of the Compton edge, which is the maximum electron recoil energy due to Compton scattering. The Compton edge determination in plastic scintillators has been a subject of intense research. We adopted a simple method introduced in [3]. This method was used for the determination of the light output of the recoil protons which had the maximum energy transfer from the incident mono-energetic neutrons. The method has the following steps: First the experimental spectrum is smoothed with a 2$^{nd}$ order low pass filter and secondly the derivative of the smoothed spectrum is calculated. The derivative minimum in the region of interest takes the shape of a Gaussian function. A Gaussian is fitted to the shape and the resulting mean value is taken as the channel which corresponds to the Compton edge, while the standard deviation is taken as the experimental resolution. The justification for the use of this method in Compton edge determination has been presented in [4]. The application of this method to the determination of the Compton edge of $^{137}$Cs can be seen in Figure 3(a),(b). The calibration line can be seen in Figure 3(c).

## 3.2 From electron energy deposition to proton energy depostion

Optical photons are produced across the path of the recoil nucleus due to de-excitation of the excited molecules. Some of the produced optical photons are absorbed by the residual excited molecules in the vicinity of the de-excited one and consequently the light output signal is reduced. This phenomenon, known as Birks law, depends on the density of the excited molecules due to recoil nucleus - molecules interaction and consequently due to the LET of the recoil nucleus. Consequently the deposited energy from protons and other ions results to lower light output compared to the same deposited energy by electrons. Light output curves of the scintillator EJ299-33 for protons and heavy ions have been measured experimentally and parameterized in [5]. A quadratic relation: $L(E) = C_0 + C_1 \cdot E + C_2 \cdot E^2$ gives the light output $L(E)$ as a function of the proton deposited energy in $MeV_{ee}$, while the coefficients $C_i$ are given in Table 1 of [5]. This relation was inverted and the deposited energy of the protons was deduced from the measured light output. Measurements with a non-shielded $^{252}$Cf source were performed. The activity of the $^{252}$Cf source was 74 kBq at Oct 12, 2010 and it is traceable to NIST standards. The $^{252}$Cf half-life is 2.645y and its spontaneous fission branching fraction is 3.096% with an average number of neutrons per fission equal to 3.768. The remaining percentage (96.904%) disintegrates in alpha particles. The measurements of energy depositions in the device from neutrons emanating from it are shown in Figure 4(a). They are compared to simulated energy deposition spectra generated using GEANT4 [6] with the QGSP_BERT_HP, physics list. The spectrum used for the source simulation is the ISO unmoderated $^{252}$Cf spectrum taken from [7].

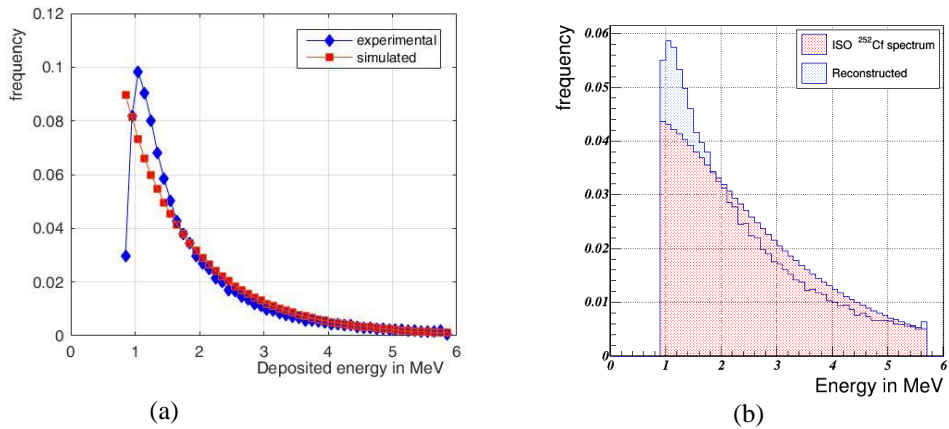

(a)                      (b)

**Figure 4.** (a) The experimental energy deposition spectrum from the $^{252}$Cf neutron source compared to the simulated one. (b) The reconstructed neutron energy spectrum.

## 3.3 $^{252}$Cf neutron spectra reconstruction

The neutron detection is based on the elastic scattering of neutrons on the light nuclei of the plastic scintillator and mainly on the hydrogen's protons. The scattering interaction transfers a portion of the neutron kinetic energy to the target nucleus, resulting in a recoil nucleus. For a monoenergetic neutron beam with kinetic energy equal to $E_n$, the spectrum of the energy transferred to the recoil protons and ions $E_d$ is a continuous function $R(E_d;E_n)$ extended in the energy range from $0$ to $E_n$. Some rare events with energy higher than $E_n$ could be detected due to nuclear reactions. In case of a continuous neutron spectrum $S_n(E_n)$ the spectrum of the transferred energy $S_t(E_d)$ is the convolution of the incident neutron spectrum with the response function $R(E_d;E_n)$.

$$S_t(E_d) = \int_0^{E_n} R(E_d; E_n) \cdot S_n(E_n) \cdot dE_n$$

The response functions $R(E_d;E_n)$ for various $E_n$ in the energy range of 0.1 to 100 MeV were determined using Monte Carlo calculations. The simulated spectra of energy depositions by mono-energetic neutron beams are used in order to infer the spectrum of $^{252}$Cf from the measured energy deposition spectrum. The idea is to select one after the other the energy bins and suppose that their entries come from recoil protons having the whole kinetic energy of the neutron and from recoil protons that have part of the energy of neutrons with energy higher than that of the energy bin. The fraction of events which result from neutrons with higher energy is estimated with the help of Monte Carlo and subtracted by the events of the energy bin. The reconstructed neutron energy spectrum is shown in Figure 4(b). One possible reason for the higher frequency of events below 1.5 MeV compared to the simulation results is the contamination of the neutron sample by gammas because of the lower discrimination capability of our system at low energies, where the neutron and the gamma branch in a scatter plot like that of Figure 2(b) converge. Furthermore one can see that measurements for energies up to 5.7 MeV have been performed. The reason for this is that the present version of the device saturates for higher energy depositions. The reduction of gain of the device in order to accommodate higher signals will not affect the relative separation between signal and noise seen in Figure 2(b), because this separation is the result of the good optical contact between the plastic scintillator and the SiPM and the good isolation from ambient light. With a previous version of the device with the same electronics but without good optical contact and isolation the gain was 12 times lower and signal from a 23 MeV monoenergetic neutron beam was collected, however the noise from dark counts was intermingled in the signal. A second source of possible error is that the Birks law parameterization used results in zero light output for 500 keV protons, which obviously is not the case, although the light output for protons with energy 500 keV or less is very low.

### 3.4 Efficiency calculation

The efficiency calculation was performed with the aid of Monte Carlo simulations. The device was placed in the center of a sphere with radius of 1 m. The neutrons were emanating from random positions on the interior surface of the sphere and they had direction with uniform distribution as far as it concerns the θ and φ directional angles. The θ angle was the polar angle with respect to the radius of the sphere at the point of the neutron production and the φ angle was the azimuthal one. The limits for the φ angle were [0,360°), and the limits of the θ angle were [0, 0.3473°). In this way the sphere with minimum radius enclosing the plastic scintillator was covered. The results are presented in Figure 5, where the number of neutrons measured has been divided by the impinging neutron fluence. So, in principle, neutron fluence as a function of energy can be deduced by the device measurements.

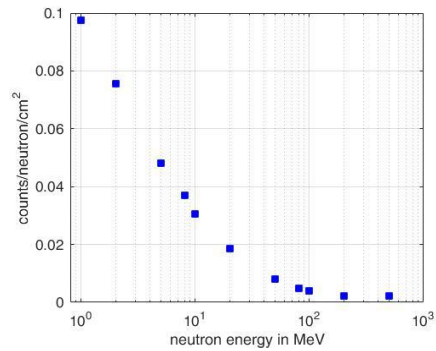

**Figure 5.** Simulated efficiency of the neutron detector

## 4. Conclusions

A miniature spectrometer for fast neutrons has been developed. It is part of the MIDAS device, which will be used for measuring the radiation field parameters near to an astronaut.

The device has been characterized and it has proven its capability to measure neutron energy depositions in the energy range from 1 MeV up to 5.7 MeV. The most important part of the fast neutron energy spectrum from the radiological point of view is from 0.1 up to 20 MeV.

The neutron flux as a function of energy can be deduced from the device measurements.

It is well known that the output current of the SiPM saturates when the number of photons impinging on it becomes comparable to the number of its micro-cells. In this case the number of fired pixels is not the same with the number of impinging photons. Due to the high gain of the present system, even with the hypothesis that all the created photons within the scintillator are impinging on the SiPM, the highest adc channel corresponds to about 4500 photons, which is well below the 18980 micro-cells of the SiPM. The extension of the measuring capability to higher energies can be achieved by changing the gain parameters of the electronic circuit. With a previous version of the device with the same electronics gain but poor optical contact signal from 23 MeV neutrons was recorded. However, as the energy goes higher, the effect of SiPM saturation has to be taken into account.

The extension of the measuring capability to lower proton energies, requires the Monte Carlo investigation of the scintillator light output for these energies, as the device can distinguish the $^{57}$Co Compton (40 keV) or even the Compton edge of $^{241}$Am (11,25 keV).

## Acknowledgments

This work is funded by the European Space Agency Contract 4000119598/17/NL/LF for the development of a highly miniaturized ASIC radiation detector.